\documentclass[12pt,a4paper]{article} 
\usepackage[left=3cm, right=3cm, top=3cm, bottom=3cm]{geometry} 

\usepackage{graphicx}
\usepackage{}
\usepackage{amssymb}
\usepackage{bm}
\usepackage{enumerate}
\usepackage{ upgreek }
\usepackage{amsmath}
\usepackage{algorithm}
\usepackage{algorithmic}
\usepackage{blkarray}
\usepackage{color}

\usepackage[dvipsnames, svgnames, x11names]{xcolor} 

\usepackage{setspace} 
\usepackage{titletoc}
\usepackage{titlesec} 
\titlelabel{\thetitle.\quad}
\usepackage{indentfirst} 
\usepackage{dcolumn}
\usepackage{rotating} 
\usepackage{longtable}

\usepackage[symbol]{footmisc}

\usepackage{authblk} 

\title{\Large\textbf{Design Selection for Two-Level Multi-Stratum Factorial Experiments Based on  Swarm Intelligence Optimization}}
\author[xie]{\normalsize\textbf{Xie-Yu Li$^{1}$, Wei-Yang Yu$^{2}$, and Ming-Chung Chang$^{2}$}\thanks{Correspnding to: Ming-Chung Chang\par \ \, E-mail: mcchang0131@gmail.com}
	\\
	$^{1}$Graduate Institute of Statistics, National Central University, Taiwan\\
	$^{2}$Institute of Statistical Science, Academia Sinica, Taiwan
}

\date{}

\begin{document}
	\maketitle
	{
	\linespread{1.5} \selectfont
	\subsection*{\centering \normalsize ABSTRACT}	
		For unstructured experimental units, the minimum aberration due to Fries and Hunter (1980) \cite{MA} is a popular criterion for choosing regular fractional factorial designs. Following which, many related studies have focused on multi-stratum factorial designs, in which multiple error terms arise from the complicated structures of experimental units. Chang and Cheng (2018) \cite{Chang2018} proposed a Bayesian-inspired aberration criterion for selecting multi-stratum factorial designs, which can be considered as a generalized version of that in  Fries and Hunter (1980)  \cite{MA}. However, they did not propose algorithms for searching for minimum aberration designs. The particle swarm optimization (PSO) algorithm is a popular optimization method that has been widely used in various applications. In this paper, we propose a new version of the PSO to select regular as well as nonregular multi-stratum designs. To select regular ones, we treat defining words as particles in the PSO and link the PSO with design key matrices. For nonregular multi-stratum designs, we treat treatment combinations as particles in the PSO. Several numerical illustrations are provided.
		\\
		
		\noindent
		Key words and phrases: Minimum aberration, Design key, Block design, Split-plot design, Strip-plot design.

	
	\section{Introduction}
	Identifying important factors is crucial for improving an industrial process.
	Multi-stratum factorial designs are an effective methodology for this purpose, especially for two-level designs owing to their run size economy.
	Industrial processes are often complex and have multiple stages, yielding complicated structures of experimental units, where split-plot designs and strip-plot designs are two common experimental plans. The structures of experimental units obtained by nesting and crossing operations are called \emph{simple block structures}. Many real applications involve experimental units with simple block structures, such as block split-plot experiments and row-column experiments.
	Evaluating and selecting multi-stratum designs have been an important research problem for decades.
	
	Minimum aberration is commonly used for assessing and selecting fractional factorial designs with unstructured units. Fries and Hunter (1980) \cite{MA} proposed the first aberration criterion for selecting two-level regular factorial designs with unstructured experimental units. When experiments become more complex, the minimum aberration is modified to other forms such as $W_{1}$  and $W_2$ in Cheng and Wu (2002) \cite{9} for blocked factorial designs. It is not easy to statistically argue an aberration criterion for complex structures of experimental units. Chang and Cheng (2018) \cite{Chang2018} proposed a Bayesian-inspired minimum aberration criterion based on the D-optimality for multi-stratum factorial designs, which can be used for \emph{orthogonal block structures} defined in Speed and Bailey (1982) \cite{orthogonal block structure}. 
	
	This paper aims at the construction of minimum aberration multi-stratum designs with simple block structures. We utilize the aberration criterion in Chang and Cheng (2018) \cite{Chang2018} to assess designs. To avoid exhaustive search, we adopt design keys due to H. D. Patterson (1976) \cite{Patterson}. Cheng and Tsai (2013) \cite{Designkey} proposed several templates of design keys, which can systematically eliminate many unnecessary searches. The advantage of using design keys can be seen more clearly for experiments with simple block structures.
	To efficiently find desirable design keys, we use the \textit{swarm intelligence-based} (SIB) algorithm in Phoa (2016) \cite{SIB} with defining words as \emph{particles}.
	SIB is a discrete version of the particle swarm optimization (PSO) algorithm in Kennedy
	and Eberhart (1995) \cite{11} in the sense that it can handle categorical inputs. There are two advantages of using PSO. First, the PSO algorithm requires few assumptions and supports
	multi-dimensional search. It has been widely used in many fields and achieved
	good results. Second, using defining words as particles and adopting the
	notion of design key can yield more efficient PSO searches. 
	
	Some preliminary materials, including simple block structures, strata and aberration criteria are presented in Section 2. Our algorithms are proposed in Section 3. Numerical illustrations under several settings are given in Section 4. We conclude our work in Section 5. The details of design keys, the PSO algorithm and the SIB algorithm are deferred to Appendix.

	\section{Preliminaries}
	We introduce the notion of simple block structures and the aberration criterion in Chang and Cheng (2018) \cite{Chang2018} in this section. The former is presented following the lines of Cheng (2014) \cite{10}, while the latter follows Chang and Cheng (2018) \cite{Chang2018}.

	\subsection{Block structure}
	Denote $\bm{\Omega}$ the set of
	$N$ experiment units. A \emph{unit factor} $\mathcal{F}$ with $n_{\mathcal{F}}$ levels can be considered as a
	partition of $\bm{\Omega}$ into $n_{\mathcal{F}}$ disjoint nonempty subsets with each consisting of the
	experiment units that have the same level of $\mathcal{F}$. These $n_{\mathcal{F}}$ subsets of $\bm{\Omega}$ are
	called $\mathcal{F}$-classes. If $N$ is a multiple of $n_{\mathcal{F}}$ and all the $\mathcal{F}$-classes have the same number of experiment
	units (i.e., $N/n_{\mathcal{F}}$), then $\mathcal{F}$ is called a \emph{uniform (unit) factor}. Two special
	unit factors are called \emph{universal} factor and \emph{equality} factor, denoted  by $\mathcal{U}$ and $\mathcal{E}$. The $\mathcal{U}$ is the coarsest among all the unit factors and has $N$ experimental units in one single class. On the other hand, $\mathcal{E}$ is the finest
	unit factor, each class consisting of one unit. In general, the notation
	$\mathfrak{B}$ denotes a set of all unit factors on the same $\bm{\Omega}$. The block structure
	$\mathfrak{B}=\{\mathcal{U},\mathcal{E}\}$ corresponds to unstructured units.
	
	Two unit factors $\mathcal{F}_1$ and $\mathcal{F}_2$ are said to be equivalent, denoted by $\mathcal{F}_1\equiv \mathcal{F}_2$, if they induce the same partition of the experimental units. If $\mathcal{F}_1$ and $\mathcal{F}_2$ are not equivalent and each $\mathcal{F}_1$-class is contained in some $\mathcal{F}_2$-class, which
	implies that any two units taking the same level of $\mathcal{F}_1$ also have the same
	level of $\mathcal{F}_2$, then we say that $\mathcal{F}_1$ is \textit{finer} than $\mathcal{F}_2$, or $\mathcal{F}_1$ is \textit{nested} in $\mathcal{F}_2$. We
	write $\mathcal{F}_1\prec\mathcal{F}_2$ if $\mathcal{F}_1$ is finer than $\mathcal{F}_2$. We also write $\mathcal{F}_1\preceq\mathcal{F}_2$ if $\mathcal{F}_1\prec\mathcal{F}_2$ or
	$\mathcal{F}_1\equiv\mathcal{F}_2$. For two unit factors $\mathcal{F}_1,\mathcal{F}_2$, denote the \textit{supremum} of $\mathcal{F}_1$ and $\mathcal{F}_2$
	by $\mathcal{F}_1\vee\mathcal{F}_2$, and \textit{infimum} of $\mathcal{F}_1$ and $\mathcal{F}_2$ by $\mathcal{F}_1\wedge\mathcal{F}_2$. The $\mathcal{F}_1\vee\mathcal{F}_2$ is the finest
	unit factor $\mathcal{G}$ satisfying $\mathcal{F}_1\preceq\mathcal{G}$ and $\mathcal{F}_2\preceq\mathcal{G}$, and $\mathcal{F}_1\wedge\mathcal{F}_2$ is the coarsest
	unit factor $\mathcal{G}^*$ satisfying $\mathcal{G}^*\preceq\mathcal{F}_1$ and $\mathcal{G}^*\preceq\mathcal{F}_2$.
	
	The definitions of \textit{crossing} and \textit{nesting} operators are essential to simple block structures and are given as follows. For any two set $\bm{\Omega}_1$ and $\bm{\Omega}_2$, let $\bm{\Omega}_1\times\bm{\Omega}_2=\{(w_1,w_2) : w_1 \in \bm{\Omega}_1,\ w_2\in
	\bm{\Omega}_2\}$. Suppose $\mathcal{F}_1$ is a factor on $\bm{\Omega}_1$ and $\mathcal{F}_2$ is a factor on $\bm{\Omega}_2$. We define
	$\mathcal{F}_1\times\mathcal{F}_2$ as the factor on $\bm{\Omega}_1\times\bm{\Omega}_2$ such that $(v_1, v_2)$ and $(w_1,w_2)$ are
	in the same ($\mathcal{F}_1\times\mathcal{F}_2$)-class if and only if $v_1$ and $w_1$ are in the same $\mathcal{F}_1$-
	class, $v_2$ and $w_2$ are in the same $\mathcal{F}_2$-class. In other words, $(\mathcal{F}_1\times\mathcal{F}_2)$ is
	a partition of $\bm{\Omega}_1\times\bm{\Omega}_2$ according to the levels of $\mathcal{F}_1$ and $\mathcal{F}_2$ simultaneously. We give the following three definitions for later use.\\
	
	\textbf{Definition 1}\quad Given a block structure $\mathfrak{B}_1$ on $\bm{\Omega}_1$  and a block structure
	$\mathfrak{B}_2$ on $\bm{\Omega}_2$, $\mathfrak{B}_1\times\mathfrak{B}_2$ is defined as the block structure on $\bm{\Omega}_1\times\bm{\Omega}_2$ that consists
	of all the factors $\{\mathcal{F}_1\times\mathcal{F}_2:\mathcal{F}_1\in\mathfrak{B}_1,\ \mathcal{F}_2\in\mathfrak{B}_2\}$.\\
	
	\textbf{Definition 2}\quad Given a block structure $\mathfrak{B}_1$ on $\bm{\Omega}_1$ and a block structure
	$\mathfrak{B}_2$ on $\bm{\Omega}_2$, $\mathfrak{B}_1/\mathfrak{B}_2$ is the block structure on $\bm{\Omega}_1\times\bm{\Omega}_2$ consisting of the
	factors in $\{\mathcal{F}_1\times\mathcal{U}_2:\mathcal{F}_1\in\mathfrak{B}_1,\ \mathcal{F}_1\neq\mathcal{E}_1\}\cup\{\mathcal{E}_1\times\mathcal{F}_2:\mathcal{F}_2\in\mathfrak{B}_2\}$, where $\mathcal{E}_1$ is
	the equality factor on $\bm{\Omega}_1$ and $\mathcal{U}_2$ is the universal factor on $\bm{\Omega}_2$.\\
	
	\textbf{Definition 3}\quad The block structure $\mathfrak{B}=\{\mathcal{U},\mathcal{E}\}$ of unstructured units
	is a simple block structure. If $\mathfrak{B}_1$ and $\mathfrak{B}_2$ are simple block structures, then
	$\mathfrak{B}_1\times\mathfrak{B}_2$ and $\mathfrak{B}_1/\mathfrak{B}_2$ are also simple block structures.\\
	
	Definition 3 gives a formal definition of simple block structures, which include block designs, split-plot designs, and strip-plot designs. By introducing the notion of \emph{orthogonality}, one can define a more general block structure referred to as orthogonal block structures.
	Two unit factors $\mathcal{F}_1$ and $\mathcal{F}_2$ are said to be \textit{orthogonal} if they have proportional frequencies
	in each $(\mathcal{F}_1\vee\mathcal{F}_2)$-class $\Upgamma$. This means that for each $(\mathcal{F}_1\vee\mathcal{F}_2)$-class $\Upgamma$, if
	both the $i$th $\mathcal{F}_1$-class and the $j$th $\mathcal{F}_2$-class are contained in $\Upgamma$, then\begin{align*}
	n_{ij}=\frac{n_{i+}n_{+j}}{|\Upgamma|},
	\end{align*}
	where $n_{ij}$ is the number of units in the intersection of the $i$th-class of $\mathcal{F}_1$ and the $j$th-class of $\mathcal{F}_2$, $n_{i+}=\sum_{j=1}^{n_{\mathcal{F}_2}}n_{ij}$, $n_{+j}=\sum_{i=1}^{n_{\mathcal{F}_1}}n_{ij}$ and $|\Upgamma|$ is the number of units
	in $\Upgamma$. The definition of orthogonal block structures is given below.\\
	
	\textbf{Definition 4}\quad A block structure $\mathfrak{B}$ consisting of nonequivalent factors
	is called an \textit{orthogonal block structure} if the following conditions hold:
	\begin{enumerate}[(O.1)]
		\item All the unit factors in $\mathfrak{B}$ are uniform and pairwise orthogonal; 
		\item $\mathcal{E}\in\mathfrak{B},\ \mathcal{U}\in\mathfrak{B}$; 
		\item If two unit factors $\mathcal{F},\ \mathcal{G}\in\mathfrak{B}$, then $\mathcal{F}\vee\mathcal{G}\in\mathfrak{B}$ and $\mathcal{F}\wedge\mathcal{G}\in\mathfrak{B}$.
	\end{enumerate}
	
	\subsection{Model and aberration criterion}
	We consider two-level treatment factors in this paper. Denote each treatment
	combination by $\mathbf{x} = (x_1,\dots,x_n)^T$, where $x_i=0$ or $1$ is the level of $i$th treatment factor. Let $\bm{\alpha}$ be a vector of the $2^n$ treatment effects $\alpha(\mathbf{x})$s under a complete factorial design. It follows that $\bm{\alpha} =\mathbf{P}\bm{\beta}$, where $\mathbf{P}$ is a $2^n \times 2^n$ model matrix for a complete factorial experiment
	and $\bm{\beta} = (\beta_{S_1},\dots,\beta_{S_{2^n}})^T$ is the vector of factorial effects $\beta_{S}$, one for each $S\subseteq\{1,\dots,n\}$. For example, for $S = \{i\},\ 1 \leq i \leq n$,\ $\beta_S$ is a main effect contrast of factor $i$, and for $S=\{i,j\}$, $1\leq i,j\leq n$, $\beta_S$ is a two-factor interaction effect of factor $i$ and factor $j$.
	
	Under the Bayesian framework, $\bm{\beta}$ is treated as a random vector. Specifying
	the distribution of $\bm{\beta}$ is a crucial step. Chang and Cheng (2018) \cite{Chang2018} regards $\bm{\alpha}$ as a realization
	of a Gaussian process. They assumed $\bm{\alpha}\sim N(\bm{0},\sigma^2\mathbf{R})$ and induced the distribution of $\bm{\beta}=\frac{1}{2^n}\mathbf{P}^{T}\bm{\alpha}$. 
	Here $\mathbf{R}$ is the correlation matrix of $\pmb{\alpha}$ at the full factorial design space and is induced by the correlation function in the Gaussian process.
	Let $\mathbf{y} = (y_1,\dots,y_N)^T$ be the responses under an $N$-run fractional factorial
	design. Suppose the $N$ experimental units have an orthogonal block structure
	$\mathfrak{B} = \{\mathcal{F}_0,\mathcal{F}_1,\dots,\mathcal{F}_m\}$, where $\mathcal{F}_0=\mathcal{U}$ and $\mathcal{F}_m = \mathcal{E}$. We adopt the mixed-effect
	model:
	\begin{align*}
	\mathbf{y}=\mathbf{X}_{T}\bm{\alpha}+\sum_{\mathcal{F}\in\mathfrak{B}}\mathbf{X}_{\mathcal{F}}\bm{\gamma}^{\mathcal{F}}
	\end{align*}
	with $\bm{\alpha}=\mathbf{P}\bm{\beta}$. Thus
	\begin{align*}
	\mathbf{y}=\mathbf{X}_{T}\mathbf{P}\bm{\beta}+\sum_{\mathcal{F}\in\mathfrak{B}}\mathbf{X}_{\mathcal{F}}\bm{\gamma}^{\mathcal{F}}
	\end{align*}
	where $\mathbf{X}_{T}$ is an $N\times 2^n$ unit-treatment incidence matrix, $\mathbf{X}_{\mathcal{F}}$ is an $N\times n_{\mathcal{F}}$ $(0,1)$-matrix in which the $(i, j)$th entry is equal to 1 if and only if the $i$th unit belong to the $j$th $\mathcal{F}$-class and $\bm{\gamma}^{\mathcal{F}_i}=(\gamma_1^{\mathcal{F}_i},\dots,\gamma_{n_{\mathcal{F}_i}}^{\mathcal{F}_i})$, where $\gamma_j^{\mathcal{F}_i}$ is the effect of $j$th level of unit factor $\mathcal{F}_i$. We assume that $\gamma_{j}^{\mathcal{F}_i}$s independently follows $N(0,\sigma^2_{\mathcal{F}_i})$ for all $j$th-classes in $\mathcal{F}_i$ and each $\bm{\gamma}^{\mathcal{F}}$ of $\mathcal{F}\in\mathfrak{B}$ are independent for $\bm{\alpha}$.
	Under this model, we have
	\begin{align*}
	\mathbf{y}|\bm{\beta}\sim N(\mathbf{X}_{T}\mathbf{P}\bm{\beta},\sum_{i=0}^{m}\sigma^2_{\mathcal{F}_{i}}\mathbf{X}_{\mathcal{F}_i}\mathbf{X}_{\mathcal{F}_i}^T).
	\end{align*}
	
	Let $\mathbf{V}=\sum_{i=0}^{m}\sigma^2_{\mathcal{F}_i}\mathbf{X}_{\mathcal{F}_i}\mathbf{X}_{\mathcal{F}_i}^T$ and $\mathbf{U}=\mathbf{X}_{T}\mathbf{P}$. Then $\mathbf{U}$ is the full model matrix under the $N$-run design. Each column of $\mathbf{U}$, except for the column of ones, corresponds to a
	factorial effect. When the block structure satisfies (O.1),
	(O.2), (O.3), $\mathbf{V}$ has $m+1$ eigenspaces $W_{\mathcal{F}_0},\dots,W_{\mathcal{F}_m}$, with one eigenspace
	associated with each of $m + 1$ unit factors. For instance, $W_{\mathcal{F}_0} =W_{\mathcal{U}}$ is the
	one-dimensional space consisting of all the vectors with constant entries,
	and each other eigenvector defines a unit contrast. Let the eigenvalues
	be $\xi_{\mathcal{F}_0},\dots,\xi_{\mathcal{F}_m}$. Therefore $\text{Var}(\mathbf{c}^T\mathbf{y}|\bm{\beta}) = ||\mathbf{c}||^2\xi_{\mathcal{F}_i}$ for each $\mathbf{c}\in W_{\mathcal{F}_i}.$ The
	eigenspaces $W_{\mathcal{F}_0},\dots,W_{\mathcal{F}_m}$ and eigenvalues $\xi_{\mathcal{F}_0},\dots,\xi_{\mathcal{F}_m}$ are called \textit{strata} and
	\textit{stratum variances}, respectively, and
	\begin{align*}
	\xi_{\mathcal{F}}=\sum_{\mathcal{G}\in\mathfrak{B}:\mathcal{G}\preceq\mathcal{F}}\frac{N}{n_{\mathcal{G}}}\sigma^2_{\mathcal{G}}.
	\end{align*}
	In addition, the case where $\bm{\gamma}^{\mathcal{F}_i} = (\gamma_1^{\mathcal{F}_i},\dots,\gamma_{n_{\mathcal{F}_i}}^{\mathcal{F}_i})$ are fixed effects can be
	treated by letting $\sigma^2_{\mathcal{F}_i}=\infty$, which leads to $\xi_{\mathcal{F}_j}=\infty$ if $\mathcal{F}_i\preceq\mathcal{F}_j$.
	
	For each nonempty $S\subseteq\{1,\dots,n\},$ let $\mathbf{u}_S$ be the column of   $ \mathbf{U}$ corresponding to $\beta_{S}$. For multi-stratum designs satisfying (O.1), (O.2), (O.3), we can define the generalized word counts in various strata as follows:
	for $k = 1,\dots, n$ and $i = 0,\dots,m$,
	\begin{align}
	B_{k,i}=\frac{1}{N}\sum_{S:|S|=k}||\mathbf{P}_{W_{\mathcal{F}_i}}\mathbf{u}_S||^2.\label{Bki}
	\end{align}
	Note that (\ref{Bki}) can be used for both regular and nonregular designs.
	Based on (\ref{Bki}), Chang and Cheng (2018) \cite{Chang2018} proposed an aberration criterion which is to sequentially minimize the following wordlength pattern:
	\begin{align}
	W(d)=\sum_{i=0}^{m-1}\left(\frac{1}{\xi_{\mathcal{F}_m}}-\frac{1}{\xi_{\mathcal{F}_i}}\right)\left(B_{1,i}(d),\dots,B_{n,i}(d)\right).\label{Wd}
	\end{align}
	They also provided a necessary and sufficient condition for
	a design to have minimum aberration for all feasible stratum
	variances. Here, $\bm{\xi}$ is said to be feasible if $\mathcal{F}_i\prec\mathcal{F}_j\Rightarrow\xi_{\mathcal{F}_i}\leq\xi_{\mathcal{F}_j}$. This condition is given in Theorem 2 as follows.\\
	
	\textbf{Theorem 2}\quad Suppose $\mathfrak{B}$ is a block structure satisfying (O.1), (O.2), (O.3). Then a necessary and sufficient condition for a design to sequentially minimize
	$W(d)$ for all feasible $\bm{\xi}$ is that it sequentially minimizes
	\begin{align*}
	W_{\mathcal{G}}(d)=\sum_{i:\mathcal{F}_i\in\mathcal{G}}\left(B_{1,i}(d),\dots,B_{n,i}(d)\right)
	\end{align*}
	for all subsets $\mathcal{G}$ of $\mathfrak{B}\setminus\{\mathcal{F}_m\}$ such that
	\begin{align}
	\mathcal{F}\in\mathcal{G},\quad\mathcal{F}'\in\mathfrak{B},\quad\text{and}\quad \mathcal{F}\prec\mathcal{F}'\Rightarrow\mathcal{F}'\in\mathcal{G}.\label{G}
	\end{align}
	Chang and Cheng (2018) \cite{Chang2018} showed that $W_{\mathcal{G}}(d)$ serves as an aberration criterion for the block structure $\mathcal{G}\cup \{\mathcal{E}\}$ with fixed unit effects.

	\section{Multi-Stratum Design Construction}
	We propose to construct minimum aberration designs with respect to $W_{\mathcal{G}}(d)$ in Theorem 2 by using design keys and the SIB algorithm. We first construct design keys and then adopt the SIB to obtain minimum aberration designs. This two-step procedure is referred to as the \emph{design-key step} and \emph{SIB step} in this paper.
	
	\subsection{Regular factorial design}
	Here we consider regular designs with simple block structures. Several corresponding design key templates have been provided in Cheng and Tsai (2013) \cite{Designkey}.
	
	\subsubsection*{Design-key step}
	Consider $2^{n}$ complete factorial designs divided into $2^b$ blocks of equal size. The block structure is $\mathfrak{B}=\{\mathcal{U},\mathcal{B},\mathcal{E}\}$, where $\mathcal{E}\prec\mathcal{B}\prec\mathcal{U}$. We set $n = 5,  b = 3$. The five treatment factors are denoted by $A,B,C,D,E$.
	We need to choose three independent block defining words to divide the $2^5$ complete factorial design into 8 blocks of size 4. In this case, a design key matrix provided in Cheng (2014) \cite{10} is the matrix $\mathbf{K}$ given below, where $\ast$ can be inserted by 0 or 1. The three independent block defining words can be determined from the inverse of $\mathbf{K}$:
	\[\small
	\mathbf{K}=
	\begin{blockarray}{cccccc}
	\mathcal{E}_1 & \mathcal{E}_2  & \mathcal{B}_1 & \mathcal{B}_2  & \mathcal{B}_3 \\
	\begin{block}{[ccccc]c}
	1 & 0 & 0 & 0 & 0 & A \\
	0 & 1 & 0 & 0 & 0 & B \\
	* & * & 1 & 0 & 0 & C \\ 
	* & * & 0 & 1 & 0 & D \\
	* & * & 0 & 0 & 1 & E \\
	\end{block}
	\end{blockarray}
	,\quad
	\mathbf{K}^{-1}=
	\begin{blockarray}{cccccc}
	A & B  & C & D  & E \\
	\begin{block}{[ccccc]c}
	1 & 0 & 0 & 0 & 0 & \mathcal{E}_1 \\
	0 & 1 & 0 & 0 & 0 & \mathcal{E}_2 \\
	* & * & 1 & 0 & 0 & \mathcal{B}_1 \\ 
	* & * & 0 & 1 & 0 & \mathcal{B}_2 \\
	* & * & 0 & 0 & 1 & \mathcal{B}_3 \\
	\end{block}
	\end{blockarray}.
	\]
	According to Remark 1 in Appendix, we note that $\mathbf{K}=\mathbf{K}^{-1}$.
	The resulting block defining contrast subgroup has $2^3 - 1=7$ defining words. 
	Let the third to the fifth rows in $\mathbf{K}^{-1}$ be $\mathcal{B}_\text{generators}$ as follows:
	\[\small
	\mathcal{B}_\text{generators}:
	\begin{blockarray}{ccccc}
	A & B  & C & D  & E \\
	\begin{block}{[ccccc]}
	* & * & 1 & 0 & 0  \\ 
	* & * & 0 & 1 & 0  \\
	* & * & 0 & 0 & 1  \\
	\end{block}
	\end{blockarray}
	.
	\]
	Each row of $\mathcal{B}_\text{generators}$ corresponds to one block generator. In the columns corresponding to $A$ and $B$, each of the three rows of $\mathcal{B}_\text{generators}$ is chosen from one of the 4 rows of $\mathbf{Pool_{\mathcal{B}}}$:
	\[\small
	\mathbf{Pool_{\mathcal{B}}}=
	\begin{blockarray}{cc}
	&     \\
	\begin{block}{[cc]}
	0 & 0 \\
	1 & 0 \\
	0 & 1 \\ 
	1 & 1 \\
	\end{block}
	\end{blockarray}.
	\]
	By inserting the second row $(1,0)$ of $\mathbf{Pool_{\mathcal{B}}}$ into the first row of $\mathcal{B}_\text{generators}$, for example, $\mathcal{B}_\text{generators}$ has $(1,0,1,0,0)$ as its first row, resulting in the block generator $AC$.
	A design key is obtained by choosing the rows of $\mathbf{Pool_{\mathcal{B}}}$ to be the rows of $\ast$ in $\mathcal{B}_\text{generators}$. An example is given below:
	\[\small
	\mathcal{B}_\text{generators}=
	\begin{blockarray}{ccccc}
	A & B  & C & D  & E  \\
	\begin{block}{[ccccc]}
	0 & 0 & 1 & 0 & 0 \\
	1 & 0 & 0 & 1 & 0 \\
	1 & 1 & 0 & 0 & 1 \\
	\end{block}
	\end{blockarray}
	,\]
	leading to
	\[\small
	\mathbf{K}^{-1}=
	\begin{blockarray}{cccccc}
	A & B  & C & D  & E &  \\
	\begin{block}{[ccccc]c}
	1 & 0 & 0 & 0 & 0 & \mathcal{E}_1 \\ 
	0 & 1 & 0 & 0 & 0 & \mathcal{E}_2 \\
	0 & 0 & 1 & 0 & 0 & \mathcal{B}_1 \\
	1 & 0 & 0 & 1 & 0 & \mathcal{B}_2 \\
	1 & 1 & 0 & 0 & 1 & \mathcal{B}_3 \\
	\end{block}
	\end{blockarray}
	.\]
	Based on $\mathbf{K}^{-1}$, the three independent block defining words are $C,\
	AD,\ ABE$. Under the hierarchical assumption (Wu and Hamada, 2009 \cite{Wu}), one should avoid confounding main effects with blocks. Thus, we need to remove undesirable rows from  $\mathbf{Pool_{\mathcal{B}}}$ to get a smaller one, which can also reduce the computational cost. 
	
	For a block structure $\mathfrak{B} = \{\mathcal{F}_0,\mathcal{F}_1,\dots,\mathcal{F}_m\}$ satisfying (O.1), (O.2), (O.3), we need to simultaneously consider different strata. We call the $l_i$ independent defining words that form $\mathcal{F}_i$-classes as
	$\mathcal{F}_i$-generator(s), where $n_{\mathcal{F}_i} = 2^{l_i}$, $i = 0, \dots, m$.  We denote each stratum pool by $\mathbf{Pool_{\mathcal{F}_i}}$
	and the reduced pool by $\mathbf{Pool_{\mathcal{F}_i}^*}$
	for $i = 0, \dots, m - 1$. Suppose the above blocked $2^5$ factorial design with four blocks is to be further blocked  into $2$ blocks. The  resulting block structure is $\mathfrak{B}=\{\mathcal{U},\mathcal{B},\mathcal{T},\mathcal{E}\}$ and $\mathcal{E}\prec\mathcal{T}\prec\mathcal{B}\prec\mathcal{U}$. A template design key in this case is:
	\[\small
	\mathbf{K}=
	\begin{blockarray}{cccccc}
	\mathcal{E}_1 & \mathcal{E}_2  & \mathcal{B}_1 & \mathcal{B}_2  & \mathcal{T}_1 \\
	\begin{block}{[ccccc]c}
	1 & 0 & 0 & 0 & 0 & A \\
	0 & 1 & 0 & 0 & 0 & B \\
	* & * & 1 & 0 & 0 & C \\ 
	* & * & 0 & 1 & 0 & D \\
	* & * & * & * & 1 & E \\
	\end{block}
	\end{blockarray}
	,\quad
	\mathbf{K}^{-1}=
	\begin{blockarray}{cccccc}
	A & B  & C & D  & E \\
	\begin{block}{[ccccc]c}
	1 & 0 & 0 & 0 & 0 & \mathcal{E}_1 \\
	0 & 1 & 0 & 0 & 0 & \mathcal{E}_2 \\
	* & * & 1 & 0 & 0 & \mathcal{B}_1 \\ 
	* & * & 0 & 1 & 0 & \mathcal{B}_2 \\
	* & * & * & * & 1 & \mathcal{T}_1 \\
	\end{block}
	\end{blockarray}.
	\]
	Thus we have
	\[\small
	\mathcal{B}_{ \text{generators}}:
	\begin{blockarray}{ccccc}
	A & B  & C & D  & E \\
	\begin{block}{[ccccc]}
	* & * & 1 & 0 & 0  \\ 
	* & * & 0 & 1 & 0  \\
	\end{block}
	\end{blockarray}
	,\quad
	\mathcal{T}_{ \text{generators}}:
	\begin{blockarray}{ccccc}
	A & B  & C & D  & E \\
	\begin{block}{[ccccc]}
	* & * & * & * & 1 \\
	\end{block}
	\end{blockarray}.
	\]
	The $\mathbf{Pool_{\mathcal{B}}}$ has $2^2=4$ combinations of $\ast$'s and the $\mathbf{Pool_{\mathcal{T}}}$ has $2^4=16$ combinations
	of $\ast$'s. Algorithm 1 below describes a way to
	get one ``particle'' in the SIB, involving $\mathbf{Pool}$ matrices, for complete factorial designs
	given a simple block structure.
	\begin{algorithm}[htb] 
		\caption{Design key for complete factorials} 
		\label{alg:Framwork} 
		\begin{algorithmic}[1] 
			\STATE Write down the simple block structure $\mathfrak{B}$.
			\STATE Check the number of generators in each stratum, denoted by $l_1,\dots,l_{m-1}$.
			\STATE Construct the design key based on $m-1$ $\mathbf{Pool}$ matrices or $\mathbf{Pool}^*$ matrices.
			\STATE \textbf{Repeat} $i=1,\dots, m-1$.\\
					  Randomly select $l_i$ rows in $\mathbf{Pool}_{\mathcal{F}_i}$, and obtain $l_i$ generators.
		\end{algorithmic}
	\end{algorithm}
	
	In complete factorial designs, a design key is a square matrix with $\sum_{i=1}^{m}l_i=n$. For $2^{n-l_0}$ regular fractional factorial designs, we have $\sum_{i=1}^{m}l_i=n-l_0$. In addition to constructing designs with unstructured units, design
	keys can also be used to construct regular fractional factorial designs with simple
	block structures. To generate $2^{n-l_0}$ distinct treatment combinations, the $n-l_0$ columns of $\mathbf{K}$ must be linearly independent. Hence, we can first use
	Algorithm 1 to obtain $n - l_0$ generators and then add
	$l_0$ rows for added factors. The added $l_0$ rows are needed to be linearly
	independent, which can be achieved by embedding the identity matrix of order $l_0$ denoted by $\mathbf{I}_{l_0}$.
	Details are given in Algorithm 2.
	\begin{algorithm}[htb] 
		\caption{Design key for fractional factorials} 
		\label{alg:Framwork} 
		\begin{algorithmic}[1] 
			\STATE Do Algorithm 1 with $n^* = n - l_0$ basic factors.
			\STATE Add $l_0$ rows in the bottom of the design key obtained in Algorithm 1.
			\STATE Generate $\mathbf{Pool}_{\mathcal{F}_0}$ (or $\mathbf{Pool}_{\mathcal{F}_0}^*$) matrix.
			\STATE Randomly select $l_0$ rows in $\mathbf{Pool}_{\mathcal{F}_0}$ (or $\mathbf{Pool}_{\mathcal{F}_0}^*$) matrix to get $\mathcal{U}_\text{generators}$.
			\STATE Obtain $\mathcal{U}$-generators  by horizontally combining $\mathcal{U}_\text{generators}$ and $\mathbf{I}_{l_0}$.
		\end{algorithmic}
	\end{algorithm}
	
	In the above example, the $2^{5-1}$ fractional factorial design with the block structure $\mathfrak{B}=\{\mathcal{U},\mathcal{B},\mathcal{T},\mathcal{E}\}$ has $\sum_{i=1}^{m}l_i=4$. One needs four basic
	factors $A,B,C,D$ to constitute a four-by-four matrix in $\mathbf{K}$. The fifth row of $\mathbf{K}$ corresponds to the treatment generator, referred to as $\mathcal{U}_\text{generators}$ given below.
	\[\small
	\mathbf{K}=
	\begin{blockarray}{ccccc}
	\mathcal{E}_1   & \mathcal{B}_1 & \mathcal{B}_2  & \mathcal{T}_1 &  \\
	\begin{block}{[cccc]c}
	1 & 0 & 0 & 0  & A \\
	* & 1 & 0 & 0  & B \\
	* & 0 & 1 & 0  & C \\ 
	* & * & * & 1  & D \\
	* & * & * & *  & E \\
	\end{block}
	\end{blockarray}
	,\quad
	\mathcal{U}_\text{generators}:
	\begin{blockarray}{cccccc}
	A & B  & C & D & |  & E \\
	\begin{block}{[cccccc]}
	* & * & * & * & | & \mathbf{I}_{1}\\
	\end{block}
	\end{blockarray}.
	\]

	\subsubsection*{SIB step}
	This step is to use the SIB algorithm to find the $\mathcal{F}_i$-generators that yield minimum aberration designs.
	For each $\mathcal{F}_i$ stratum, we need to select $l_i$ generators. In this paper, we treat a possible combination of the $\mathcal{F}_i$-generators in one design key as one particle in the SIB algorithm. As mentioned in Appendix, the substituting $\mathbf{q} = (q_0,\dots,q_{m-1})$ is user-defined, while here we have the restriction $0\leq q_i \leq l_i$ for $i=0,\dots,m - 1$. Let $\mathbf{X}$ be a particle and $\mathbf{X}_{-r}$ be obtained by removing the $r$th row from $\mathbf{X}$. MIX operation in the original SIB selected one row to be added to $\mathbf{X}_{-r}$ , such that the resulting X has the best minimum aberration. However, $\mathbf{X}_{-r}$ in the MIX operation in original SIB algorithm cannot be used to select the best particle among
	all candidates. We define a new $\mathbf{X}^*_r$
	to replace $\mathbf{X}_{-r}$ and modify the MIX operation.
	The $\mathbf{X}^*_r$ swaps the $r$th generator with another generator 
	coming from the following three sources:
	\begin{itemize}
		\item 	the $r$th generator in the global best particle,
		\item 	the $r$th generator in the local best particle,
		\item 	randomly select a generator in $\mathbf{Pool}^*_{\mathcal{F}}$.
	\end{itemize}

	The design key in the previous example has $l_1 = 3$, and there are three generators in a
	particle. One possible choice is to set $q_1 = 2$. The second iteration executes the MIX operation twice to construct a design (particle).
	Following the PSO in Appendix, $x_{i,n+1}$ depends on directions of global best and
	local best. Due to the three sources, we suggest a new direction of a new point in the PSO:
	\begin{align*}
	v_{i,n+1}&=v_{i,n}+C_1(x_{GB}-x_{i,n})+C_2(x_{i,LB}-x_{i,n})+C_3(x_{i,NEW}-x_{i,n}),\\
	x_{i,n+1}&=x_{i,n}+v_{i,n+1}\times\delta_t,
	\end{align*}
	where $x_{i,NEW}$ is the new position for the $i$th particle.
	Following the same spirit, we propose a new version of SIB.  Let 
	\begin{align*}
	&\mathbf{q}_{GB}= (q_{0,GB},\dots,q_{m-1,GB}),\\
	&\mathbf{q}_{LB}= (q_{0,LB},\dots,q_{m-1,LB}),\\
	&\mathbf{q}_{NEW}= (q_{0,NEW},\dots,q_{m-1,NEW}),
	\end{align*}
	respectively represent substituting $$\mathbf{q}=\{(q_{0,GB}+q_{0,LB}+q_{0,NEW}),\dots,(q_{m-1,GB}+q_{m-1,LB}+q_{m-1,NEW})\}.$$
	 All generators in the $j$th particle at the $(n + 1)$th iteration $\mathbf{X}_{j,n+1}$ are composed of:
	\begin{enumerate}
		\item $\sum_{i=0}^{m-1}q_{i,GB}$ generators from $\mathbf{X}^{GB}_{n}$ that is the global best one after $n$ iterates;
		\item $\sum_{i=0}^{m-1}q_{i,LB}$ generators from $\mathbf{X}^{LB}_{j,n}$ that is the best one in $j$th particle after $n$ iterates;
		\item $\sum_{i=0}^{m-1}q_{i,NEW}$ generators from $\mathbf{X}^{NEW}_{j,n}$ that is a new particle in $j$th particle at $n$th iteration;
		\item remaining generators in $\mathbf{X}_{j,n}$.
	\end{enumerate}

In this procedure, the two randomization mechanisms, three sources of generators, and the $r$th iteration in the MIX operation avoid the SIB algorithm
being too rigid to find global best solution.
	In practice, we fix the values of $\sum_{i=0}^{m-1}q_{i,NEW}$, $\sum_{i=0}^{m-1}q_{i,GB}$ and $\sum_{i=0}^{m-1}q_{i,LB}$, and suggest 
	$$\sum_{i=0}^{m-1}q_{i,NEW}\geq\sum_{i=0}^{m-1}q_{i,GB}\geq\sum_{i=0}^{m-1}q_{i,LB}\geq 0.$$
	The details are given in Algorithm 3.
	
	\begin{algorithm}[h] 
		\caption{ Regular multi-stratum designs} 
		\label{alg:Framwork} 
		\begin{algorithmic}[1] 
			\STATE Decide the number of particles $S$, total number of iterations $T$ and the number of
			substituting $\mathbf{q}_{GB}$, $\mathbf{q}_{LB}$, $\mathbf{q}_{NEW}$.
			\STATE Set $t=1$.
			\STATE Generate $j = 1, \dots, S$ particles by Algorithms 1 and 2, and compare them via $W_{\mathcal{G}}(d)$.
			\STATE Each particle is $\mathbf{X}^{LB}_{j,1}$ in which the best one is $\mathbf{X}^{GB}_{1}$ for all $j$.
			\STATE Set $t=2$.
			\STATE \textbf{while} $t\leq T$ \textbf{do}
			\STATE \hspace{\algorithmicindent} \textbf{MIX operation}, get a mixing particle $\mathbf{X}_{j,t}$ for all $j$.
			\STATE \hspace{\algorithmicindent} \textbf{MOVE operation}, $\mathbf{X}^{LB}	_{j,t}$ is the better of $(\mathbf{X}_{j,t},\ \mathbf{X}^{LB}_{j,t-1})$ and then the best\\\hspace{0.55cm}$\mathbf{X}^{LB}_{j,t}$ is $\mathbf{X}^{GB}_{t}$ for all $j$.
			\STATE \hspace{\algorithmicindent} $t=t+1$.
			\STATE \textbf{end while}
		\end{algorithmic}
	\end{algorithm}
	
	Theorem 2 induces several wordlength patterns $W_{\mathcal{G}}(d)$ for all $\mathcal{G}$ satisfying (\ref{G}). A design that has minimum aberration with respect to these $W_{\mathcal{G}}(d)$ is optimal with respect to (\ref{Wd}) for all feasible stratum variances. For example, given
	a block structure $\mathfrak{B}=\{\mathcal{U},\mathcal{F}_1,\mathcal{F}_2,\mathcal{F}_3,\mathcal{E}\}$, $\mathcal{E}\prec\mathcal{F}_2\prec\mathcal{F}_1\prec\mathcal{U}$ and $\mathcal{E}\prec\mathcal{F}_3\prec\mathcal{F}_1\prec\mathcal{U}$, we need to check $W_{\mathcal{G}}(d)$ for five situations $\{\mathcal{U}\}$, $\{\mathcal{U},\mathcal{F}_1\}$, $\{\mathcal{U},\mathcal{F}_1,\mathcal{F}_2\}$,  $\{\mathcal{U},\mathcal{F}_1,\mathcal{F}_3\}$ and  $\{\mathcal{U},\mathcal{F}_1,\mathcal{F}_2,\mathcal{F}_3,\mathcal{E}\}$.
	In this case, $\xi_{\mathcal{F}_2}$ can be larger or smaller than $\xi_{\mathcal{F}_3}$ because neither $\mathcal{F}_2$ nor $\mathcal{F}_3$ is nested in the other.
	In general, the ordering of some $\xi_{\mathcal{F}}$s are not determined due to the existence of crossing operations in the block structure.
	To deal with this situation, we propose two kinds of criteria referred to as the \textit{forward criterion} and the \textit{backward criterion}.
	The forward criterion regards the treatment generators
	causing non-estimable factorial effects as the most severe. So $\mathcal{U}$
	should be considered first. Then we sequentially consider $\mathcal{F}_{i1},\mathcal{F}_{i2},\dots,\mathcal{F}_{im}$,
	where $\xi_{i1}\geq\xi_{i2}\geq\dots\geq\xi_{im}$. If the crossing operation exists in the block structure, the stratum containing more alias
	effects needs to be considered first. Alternatively, the backward criterion follows a reverse path of the
	forward criterion. The procedure is given in Table 1.
	
	\begin{table}[h]  
		\centering  
		\caption{Settings of $\bm{\xi}$ for a chain of nesting $\mathcal{F}_i$} 
		\begin{tabular}{c|c|c}  
			\hline
			Sequential ordering & Forward criterion & Backward criterion  \\ [0.5ex]  
			\hline
			1. & $\{\xi_0=\infty,\ \xi_1=\dots=\xi_m\}$ & $\{\xi_0=\dots=\xi_{m-1}=\infty,\ \xi_m\}$  \\
			2. & $\{\xi_0=\xi_1=\infty,\ \xi_2=\dots=\xi_m\}$ & $\{\xi_0=\dots=\xi_{m-2}=\infty,\ \xi_{m-1}=\xi_m\}$  \\
			\vdots & \vdots & \vdots  \\		
			$m$. & $\{\xi_0=\xi_{m-1}=\infty,\ \xi_m\}$ & $\{\xi_0=\infty,\ \xi_{1}=\dots=\xi_m\}$ \\		
			\hline
		\end{tabular}
	\end{table}

	\subsection{Nonregular factorial design}
	Nonregular designs do not have defining relations. 
	Here we regard treatment combinations as particles in the SIB. There is only one pool matrix denoted by $\textbf{Pool}_{non}$, which is the complete factorial design:
	\begin{align*}
	\small
	\textbf{Pool}_{non}=\left[ \begin{array}{cccc}
	-1 & -1 & \dots & -1   \\
	& \dots &  &    \\
	1 & 1  & \dots & 1   \\ 
	\end{array}
	\right]_{2^n\times n}.
	\end{align*}
	We propose an algorithm, referred to as Algorithm 4, for nonregular designs with simple block structures.
	Under highly fractionated designs, this algorithm
	may not quickly converge.
	We suggest the following setting to avoid locally optimal solutions: $q_{NEW}\geq q_{GB}\geq q_{LB}\geq 0$ and $(q_{NEW}+q_{GB}+q_{LB})\leq N$.
	\begin{algorithm}[h] 
		\caption{Nonregular design} 
		\label{alg:Framwork} 
		\begin{algorithmic}[1] 
			\STATE Decide the number of particles $S$, total number of iterations $T$ and the number of
			substituting $q_{GB}$, $q_{LB}$, $q_{NEW}$ and fix incidence matrices $X_{\mathcal{F}_i},\  i = 1,\dots, m - 1$.
			\STATE Set $t=1$.
			\STATE Generate $j = 1, \dots, S$ particles by $\textbf{Pool}_{non}$ and compare them via $W_{\mathcal{G}}(d)$.
			\STATE Each particle is $\mathbf{X}^{LB}_{j,1}$ in which the best one is $\mathbf{X}^{GB}_{1}$ for all $j$.
			\STATE Set $t=2$.
			\STATE \textbf{while} $t\leq T$ \textbf{do}
			\STATE \hspace{\algorithmicindent} \textbf{MIX operation}, get a mixing particle $\mathbf{X}_{j,t}$ for all $j$.
			\STATE \hspace{\algorithmicindent} \textbf{MOVE operation}, $\mathbf{X}^{LB}_{j,t}$ is the better of $(\mathbf{X}_{j,t},\ \mathbf{X}^{LB}_{j,t-1})$ and then the best\\\hspace{0.55cm}$\mathbf{X}^{LB}_{j,t}$ is $\mathbf{X}^{GB}_{t}$ for all $j$.
			\STATE \hspace{\algorithmicindent} $t=t+1$.
			\STATE \textbf{end while}
		\end{algorithmic}
	\end{algorithm}

As mentioned in Luke (2013) \cite{Luke_2013}, the SIB algorithm does not guarantee that the solution obtained is optimal
with respect to a criterion. One may need to try several parameter settings in the SIB, such as the number of particles, to obtain the best design. In Section 4, the parameter settings reported are those leading to the best results.
	
	\section{Application}
	We first consider regular multi-stratum designs and then followed by nonregular multi-stratum designs.
	All unit effects are treated as fixed effects.
	Each example separately utilizes the forward criterion and the backward
	criterion to select minimum aberration designs.
	In this section, each $W_{\mathcal{G}}(d)$ defined in Theorem 2 is expressed by a set ``$\mathcal{G}$-MA''.
	
	\subsection{Regular design}
	\subsubsection*{Blocked fractional factional design}
	Chang(2018) \cite{Chang2018} studied regular $2^{13-8}$
	designs with 8 blocks of size 4 and gave
	three admissible designs, denoted by $d_1,\ d_2$ and $d_3$. Here $l_0=8$ is large, which
	causes serious aliasing between factorial effects. 
	
	Denote $A$ to $M$ the thirteen treatment factors. The block structure is $\mathfrak{B} = \{\mathcal{U}, \mathcal{B}, \mathcal{E}\}$ with $\mathcal{E}\prec\mathcal{B}\prec\mathcal{U}$. The subsets satisfying (\ref{G}) are $\mathcal{G}_1 = \{\mathcal{U}\}$ (unstructured units) and $\mathcal{G}_2 = \{\mathcal{U}, \mathcal{B}\}$ (fixed block effects). We have $n = 13,\ l_0 = 8,\ l_1 = 3$ and must choose eight $\mathcal{U}_\text{generators}$
	 and three $\mathcal{B}_\text{generators}$  to construct designs. The design key with the basic factors $(A,B,C,D,E)$ and the generators are:
	\begin{align*}
	\small
	\mathbf{K}=
	\begin{blockarray}{cccccc}
	\mathcal{E}_1   & \mathcal{E}_2  & \mathcal{B}_1  & \mathcal{B}_2 & \mathcal{B}_3 &   \\
	\begin{block}{[ccccc]c}
	1 & 0 & 0 & 0 & 0 & A \\
	0 & 1 & 0 & 0 & 0 & B \\
	*^b & *^b & 1 & 0 & 0 & C \\ 
	*^b & *^b & 0 & 1 & 0 & D \\
	*^b & *^b & 0 & 0 & 1 & E \\
	* & * & * & * & * & F \\
	&  & \vdots &  &  & \vdots \\
	* & * & * & * & * & M \\
	\end{block}
	\end{blockarray},\quad
	\begin{split}
	\mathcal{B}_\text{generators}:
	\begin{blockarray}{ccccc}
	A & B  & C & D & E \\
	\begin{block}{[ccccc]}
	*^b & *^b & 1 & 0 & 0  \\ 
	*^b & *^b & 0 & 1 & 0  \\
	*^b & *^b & 0 & 0 & 1  \\
	\end{block}
	\end{blockarray},              \\
	\\
	\mathcal{U}_\text{generators}:
	\begin{blockarray}{ccc}
	A\sim E & |  & F\sim M \\
	\begin{block}{[ccc]}
	\bm{*}_{8\times5} & | & \mathbf{I}_{8}  \\
	\end{block}
	\end{blockarray}.              
	\end{split}
	\end{align*}
	The corresponding $\textbf{Pool}_{\mathcal{U}}^*$ and $\textbf{Pool}_{\mathcal{B}}^*$ are given below, where $\textbf{Pool}_{\mathcal{U}}^*$ ensures designs of resolution at
	least three and $\textbf{Pool}_{\mathcal{B}}^*$ avoids treatment main effects confounding
	with each other:	
	\[\small
	\textbf{Pool}^*_{\mathcal{U}}=
	\begin{blockarray}{ccccc}
	& & & & \\
	\begin{block}{[ccccc]}
	1 & 1 & 0 & 0 & 0  \\ 
	1 & 0 & 1 & 0 & 0  \\
	&  & \vdots &  &   \\
	0 & 1 & 1 & 1 & 1  \\
	1 & 1 & 1 & 1 & 1  \\
	\end{block}
	\end{blockarray}
	,\quad
	\textbf{Pool}^*_{\mathcal{B}}=
	\begin{blockarray}{cc}
	&  \\
	\begin{block}{[cc]}
	1 & 0  \\
	0 & 1  \\
	1 & 1  \\
	\end{block}
	\end{blockarray}.
	\]	
	
	We set $\sum_{i=0}^{1}q_{i,NEW}=5,$ $\sum_{i=0}^{1}q_{i,GB}=4$, $\sum_{i=0}^{1}q_{i,LB}=1$, $S=50$ and $T=50$.
	Two aberration criteria are used in this example: $\mathcal{G}_1$-MA=$\{B_{1,0}(d),\dots,B_{n,0}(d)\}$ and $\mathcal{G}_2$-MA=$\{B_{1,0}(d)+B_{1,1}(d),\dots,B_{n,0}(d)+B_{n,1}(d)\}$.  
	\begin{table}[b]\footnotesize 
		\centering  
		\caption{Minimum aberration blocked fractional factional designs} 
		\begin{tabular}{l|l}  
			\hline
			Forward criterion, Chang and Cheng (2018), $d_2$ &    \\ [0.5ex]  
			\hline
			$\mathcal{G}_1$-MA & \{0, 0, 0, 55, 0, 96, 0, 87, 0, 16, 0, 1, 0\}  \\
			$\mathcal{G}_2$-MA & \{0, 36, 0, 365, 0, 848, 0, 651, 0, 140, 0, 7, 0\}  \\
			\hline
			Backward criterion, Chang and Cheng (2018), $d_1$ &    \\ [0.5ex]  
			\hline
			$\mathcal{G}_1$-MA & \{0, 0, 4, 39, 32, 48, 56, 39, 32, 0, 4, 1, 0\}  \\
			$\mathcal{G}_2$-MA & \{0, 22, 80, 163, 320, 452, 416, 311, 192, 70, 16, 5, 0\}  \\
			\hline
			Chang and Cheng (2018), $d_3$ &    \\ [0.5ex]  
			\hline	
			$\mathcal{G}_1$-MA & \{0, 0, 4, 38, 32, 52, 56, 33, 32, 4, 4, 0, 0\}  \\
			$\mathcal{G}_2$-MA & \{0, 30, 36, 255, 240, 452, 472, 255, 240, 30, 36, 1, 0\}  \\
			\hline		
		\end{tabular}
	\end{table}
	We obtain many isomorphic designs that are equivalent to
	the two minimum aberration designs obtained by Chang and Cheng (2018) \cite{Chang2018} through the forward criterion and the backward criterion, respectively. Denote the two designs by $d_1$ and $d_2$. Chang and Cheng (2018) \cite{Chang2018} also found another admissible design $d_3$, which is classified as locally optimal in our method. Since $d_2$ has the smallest $B_{3,0}(d) = 0$, it performs the best in the forward
	criterion, while $d_1$ has the smallest $B_{2,0}(d)+B_{2,1}(d) = 22$ and it performs the best in the backward criterion. It can be shown that the minimum aberration design is unique (up to isomorphism) if the forward
	criterion and the backward criterion lead to the same conclusion. 
	The values of $\mathcal{G}_1$-MA and $\mathcal{G}_2$-MA of the three designs are provided in Table 2.
	\subsubsection*{Blocked fractional strip-plot design}
	We consider blocked strip-plot designs, formed by nesting and crossing operations, in this example. Note that certain treatment factors must be estimated in certain strata.
	The block structure is $\mathfrak{B}=\{\mathcal{U},\mathcal{B},\mathcal{R},\mathcal{C},\mathcal{E}\}$ with $\mathcal{E}\prec\mathcal{R}\prec\mathcal{B}\prec\mathcal{U}$ and $\mathcal{E}\prec\mathcal{C}\prec\mathcal{B}\prec\mathcal{U}$. Let
	\begin{itemize}
		\item $N$: runs size,
		\item $n$: number of treatment factors,
		\item $l_1$: number of $\mathcal{B}$-generators,
		\item $n_1$: number of row treatment factors,
		\item $n_2$: number of column treatment factors,
		\item $l_{0r}$: number of $\mathcal{U}$-generators in the row design,
		\item $l_{0c}$: number of $\mathcal{U}$-generators in the column design,
		\item $r = n_1 - l_{0r}$, producing $2^r$ units in the row design,
		\item $c = n_2 - l_{0c}$, producing $2^c$ units in the column design.
	\end{itemize}
	Given that no combinations of row (column) treatment factors are replicated
	in the row (column) design, the block structure can be expressed as
	\begin{align*}
	2^{l_1}/(2^{r-l_1}\times2^{c-l_1}),\text{ with } N=2^{l_1}\times2^{r-l_1}\times2^{c-l_1}=2^{(n_1+n_2)-(l_{0r}+l_{0c}+l_1)}.
	\end{align*}
		
	Cheng (2014, p. 323) \cite{10} studied blocked strip-plot designs with the block structure
	$2^1/(2^2 \times 2^2)$ and $n_1=6$ row factors denoted by capital letter $A$ to $F$, and $n_2 = 4$ column factors denoted by $G$ to $J$. The other setting are $n = 10,\ l_1 = 1,\ r = c = 3,\ l_{0r} = 3$ and $l_{0c} = 1$. It is formed by two fractional factorial designs, a $2^{6-3}$ row design with $2^1$ blocks of size four and a $2^{4-1}$ column design with $2^1$ blocks of size four. We first write down a design key for the complete factorial of $n_1+n_2-l_{0r}-l_{0c}-l_1$ basic factors and focus on choosing $\mathcal{B}$-generator in the row design:
	\[\small
	\mathbf{K}=
	\begin{blockarray}{cccccc}
	\mathcal{R}_1 & \mathcal{R}_2  & \mathcal{C}_1 & \mathcal{C}_2  & \mathcal{B}_1 \\
	\begin{block}{[ccccc]c}
	1 & 0 & 0 & 0 & 0 & A \\
	0 & 1 & 0 & 0 & 0 & B \\
	0 & 0 & 1 & 0 & 0 & G \\ 
	0 & 0 & 0 & 1 & 0 & H \\
	*^r & *^r & 0 & 0 & 1 & C \\
	\end{block}
	\end{blockarray}
	,\quad
	\mathbf{K}^{-1}=
	\begin{blockarray}{cccccc}
	A & B  & G & H  & C \\
	\begin{block}{[ccccc]c}
	1 & 0 & 0 & 0 & 0 & \mathcal{R}_1 \\
	0 & 1 & 0 & 0 & 0 & \mathcal{R}_2 \\
	0 & 0 & 1 & 0 & 0 & \mathcal{C}_1 \\ 
	0 & 0 & 0 & 1 & 0 & \mathcal{C}_2 \\
	*^r & *^r & 0 & 0 & 1 & \mathcal{B}_1 \\
	\end{block}
	\end{blockarray}.
	\]	
	The fifth row of $\mathbf{K}^{-1}$ corresponds to the row treatment factor
	and the first to fourth rows correspond to the column treatment factors.
	The design key can be used to choose $\mathcal{B}$-generator in the column design:	
	\[\small
	\mathbf{K}=
	\begin{blockarray}{cccccc}
	\mathcal{R}_1 & \mathcal{R}_2  & \mathcal{C}_1 & \mathcal{C}_2  & \mathcal{B}_1 \\
	\begin{block}{[ccccc]c}
	1 & 0 & 0 & 0 & 0 & A \\
	0 & 1 & 0 & 0 & 0 & B \\
	0 & 0 & 1 & 0 & 0 & G \\ 
	0 & 0 & 0 & 1 & 0 & H \\
	0 & 0 & *^c & *^c & 1 & I \\
	\end{block}
	\end{blockarray}
	,\quad
	\mathbf{K}^{-1}=
	\begin{blockarray}{cccccc}
	A & B  & G & H  & I \\
	\begin{block}{[ccccc]c}
	1 & 0 & 0 & 0 & 0 & \mathcal{R}_1 \\
	0 & 1 & 0 & 0 & 0 & \mathcal{R}_2 \\
	0 & 0 & 1 & 0 & 0 & \mathcal{C}_1 \\ 
	0 & 0 & 0 & 1 & 0 & \mathcal{C}_2 \\
	0 & 0 & *^c & *^c & 1 & \mathcal{B}_1 \\
	\end{block}
	\end{blockarray}.
	\]
	
	Suppose the first $\mathcal{B}$-generator is $AC$, which divides the row design into $2^1$ blocks, and the second $\mathcal{B}$-generator is $GI$, which divides the column design into $2^1$ blocks. The total run sizes is $2^5$, which indicates the number of $\mathcal{U}$-generators is 5. There are $l_{0r} = 3$ generators involving the row design only and $l_{0c} = 1$ generator involving the column design only. The remaining
	$l_1 = 1$ $\mathcal{U}$-generator is used for 4 blocks which comes from crossing
	two blocks in the row design with two blocks in the column design. 
	Following 
	Lemma 1 in Appendix, the $\mathcal{U}$-generator is $ACGI$:
	\[\small
	l_{1}\ \mathcal{U}_\text{generator}=
	\begin{blockarray}{cccccc}
	A & B  & G & H  & C & I \\
	\begin{block}{[cccccc]}
	*^r & *^r & *^c & *^c & 1 & 1 \\
	\end{block}
	\end{blockarray}.
	\]
	 No matter which $\mathcal{B}$-generator is chosen, the design is the same up to isomorphism.

	The $l_{0r}+l_{0c}$ $\mathcal{U}$-generators involving only row treatment factors and column treatment factors separately can be obtained by:
	\[\small
	l_{0r}\ \mathcal{U}_\text{generator}=
	\begin{blockarray}{cccccc}
	A & B  & C & D  & E & F \\
	\begin{block}{[cccccc]}
	* & * & * & 1 & 0 & 0 \\
	* & * & * & 0 & 1 & 0 \\
	* & * & * & 0 & 0 & 1 \\
	\end{block}
	\end{blockarray},\quad
	l_{0c}\ \mathcal{U}_\text{generator}=
	\begin{blockarray}{cccc}
	G & H  & I & J   \\
	\begin{block}{[cccc]}
	* & * & * & 1  \\
	\end{block}
	\end{blockarray}.
	\]
	The design $d_1$ in Cheng (2014, p. 323) \cite{10} includes $l_1$ $\mathcal{U}$-generator $ACGI$, $l_{0r}$ $\mathcal{U}$
	generators in the row design $ABD,ABCE,BCF$, $l_{0c}$ $\mathcal{U}$-generators in the column
	design $GHIJ$ and $\mathcal{B}$-generator $AC$ (or $GI$). The subsets $\mathcal{G}$ satisfying (\ref{G})
	are $\mathcal{G}_1 = \{\mathcal{U}\}$, $\mathcal{G}_2 = \{\mathcal{U}, \mathcal{B}\}$, $\mathcal{G}_3 = \{\mathcal{U}, \mathcal{B},\mathcal{R}\}$, $\mathcal{G}_4 = \{\mathcal{U}, \mathcal{B}, \mathcal{C}\}$ and $\mathcal{G}_5 =
	\{\mathcal{U}, \mathcal{B},\mathcal{R}, \mathcal{C}\}$. Under the hierarchy principle, we apply the sequence of the forward criterion
	to be $\mathcal{G}_1 \rightarrow \mathcal{G}_2 \rightarrow \mathcal{G}_3 \rightarrow \mathcal{G}_4 \rightarrow \mathcal{G}_5$ if $l_{0r} > l_{0c}$, and $\mathcal{G}_1 \rightarrow \mathcal{G}_2 \rightarrow \mathcal{G}_4 \rightarrow \mathcal{G}_3 \rightarrow \mathcal{G}_5$ if $l_{0c} > l_{0r}$. Similar to the backward criterion, we have the sequences $\mathcal{G}_5 \rightarrow \mathcal{G}_4 \rightarrow \mathcal{G}_3 \rightarrow \mathcal{G}_2 \rightarrow \mathcal{G}_1$ and $\mathcal{G}_5 \rightarrow \mathcal{G}_3 \rightarrow \mathcal{G}_4 \rightarrow \mathcal{G}_2 \rightarrow \mathcal{G}_1$.
	The $\mathcal{G}_3$- and $\mathcal{G}_4$-MA have $B_{1,0}(d) + B_{1,1}(d) + B_{1,2}(d) = 6$ and $B_{1,0}(d) +
	B_{1,1}(d) + B_{1,3}(d) = 4$. It is common in the strip-plot design with $\mathcal{R}$-generators being
	$A,B$ and $\mathcal{C}$-generators being $G,H$; that is, confound row blocks and column blocks with some main effects. This block structure has $l_{0r} > l_{0c}$. The
	forward criterion has the same results in two different sequences and the
	backward criterion also exhibits the same phenomenon. A reasonable explanation
	is that blocked strip-plot designs can only change $\mathcal{U}$-generators and $\mathcal{B}$-generators. Hence assessing $\mathcal{G}_1$ and $\mathcal{G}_2$ is actually crucial.
	We note that $d_1$ (respectively, $d_2$) in Cheng (2014, p. 323) \cite{10} is equivalent to using the forward criterion (respectively, backward criterion). The values of $\mathcal{G}_i$-MA of the two designs are provided in Table 3.
	\begin{table}[!t]  
		\centering  
		\caption{Minimum aberration blocked fractional strip-plot designs} 
		\begin{tabular}{l|l}  
			\hline
			[Forward] $\mathcal{G}_1 \rightarrow \mathcal{G}_2 \rightarrow \mathcal{G}_3 \rightarrow \mathcal{G}_4 \rightarrow \mathcal{G}_5$,&    \\ [0.5ex] 
			$\mathcal{G}_1 \rightarrow \mathcal{G}_2 \rightarrow \mathcal{G}_4 \rightarrow \mathcal{G}_3 \rightarrow \mathcal{G}_5$, Cheng (2014), $d_1$&    \\ [0.5ex] 
			\hline
			$\mathcal{G}_1$-MA & \{0, 0, 4, 10, 8, 0, 4, 5, 0, 0\}  \\
			$\mathcal{G}_2$-MA & \{0, 5, 8, 10, 16, 10, 8, 5, 0, 1\}  \\
			$\mathcal{G}_3$-MA & \{6, 17, 32, 46, 52, 46, 32, 17, 6, 1\}  \\
			$\mathcal{G}_4$-MA & \{4, 9, 24, 54, 72, 54, 24, 9, 4, 1\}  \\
			$\mathcal{G}_5$-MA & \{10, 21, 48, 90, 108, 90, 48, 21, 10, 1\}  \\
			\hline
			[Backward] $\mathcal{G}_5 \rightarrow \mathcal{G}_4 \rightarrow \mathcal{G}_3 \rightarrow \mathcal{G}_2 \rightarrow \mathcal{G}_1$,&    \\ [0.5ex] 
			$\mathcal{G}_5 \rightarrow \mathcal{G}_3 \rightarrow \mathcal{G}_4 \rightarrow \mathcal{G}_2 \rightarrow \mathcal{G}_1$, Cheng (2014), $d_2$&    \\ [0.5ex] 
			\hline
			$\mathcal{G}_1$-MA & \{0, 0, 5, 6, 7, 8, 3, 1, 1, 0\}  \\
			$\mathcal{G}_2$-MA & \{0, 4, 10, 6, 14, 20, 6, 1, 2, 0\}  \\
			$\mathcal{G}_3$-MA & \{6, 16, 28, 42, 56, 56, 36, 13, 2, 0\}  \\
			$\mathcal{G}_4$-MA & \{4, 9, 24, 54, 72, 54, 24, 9, 4, 1\}  \\
			$\mathcal{G}_5$-MA & \{10, 21, 42, 90, 114, 90, 54, 21, 4, 1\}  \\
			\hline	
		\end{tabular}
	\end{table}
	
	\subsection{Nonregular design}
		\subsubsection*{Eight-run nonregular designs with unstructured units}
	For unstructured units, the block structure is $\mathfrak{B} = \{\mathcal{U},\mathcal{E}\}$. We need to check $\mathcal{G}_1 = \{\mathcal{U}\}$ when applying Theorem 2. Here we consider selecting eight-run nonregular designs, separately with six and seven treatment factors.
	
	In searching for minimum aberration eight-run nonregular factorial designs with six treatment factors, by Algorithm 4, the $\textbf{Pool}_{non}$ consists of all $2^6=64$ treatment
	combinations, and the particles come from randomly choosing 8 of them. We set $S = 100$, $q_{GB} = 2$, $q_{LB} = 2$ , $q_{NEW} = 4$ and then get the minimum aberration design $d_1^\ast$ with $\mathcal{G}_1$-$\text{MA}=\{0, 0, 4, 3, 0, 0\}$. To assess our method, we check the following orthogonal array $\mathbf{M}$ given by Cheng (2014, p. 123) \cite{10}, which is an orthogonal array of strength two, represented by OA(8,$2^6$,2):
	
	\[\tiny
	\mathbf{M}=
	\begin{blockarray}{rrrrrr}
	A & B  & C & D  & E & F \\
	\begin{block}{[rrrrrr]}
	-1 &-1& -1 &-1 &-1&-1 \\
	1& -1 &-1& 1& -1& 1 \\
	-1 &1& -1& 1& 1& 1 \\
	1 &1 &-1 &-1 &1 &-1 \\ 
	-1 &-1 &1 &-1& 1& 1 \\
	1 &-1& 1& 1& 1& -1 \\
	-1 &1& 1& 1 &-1& -1 \\
	1& 1& 1 &-1& -1& 1 \\
	\end{block}
	\end{blockarray}
	,\quad
	d_1^\ast=
	\begin{blockarray}{rrrrrr}
	A & B  & C & D  & E & F \\
	\begin{block}{[rrrrrr]}
	-1& -1& -1& 1& 1& -1 \\
	1& -1 &-1& -1& -1& -1 \\
	-1& 1& 1& -1& -1 &-1 \\ 
	-1& -1 &1& 1& -1& 1 \\
	1 &-1 &1 &-1& 1 &1 \\ 
	-1& 1& -1& -1 &1& 1 \\
	1& 1& 1& 1& 1 &-1 \\
	1& 1& -1& 1& -1& 1 \\
	\end{block}
	\end{blockarray}.
	\]
	The $\mathbf{M}$ has the same $\mathcal{G}_1$-$\text{MA}$ as $d_1^\ast$, while they have different treatment combinations. This is a case where the minimum aberration design is a regular design and hence can be obtained by Algorithm 3 instead of Algorithm 4.
	
	 Next, we search for minimum aberration eight-run nonregular factorial designs with seven treatment factors. By Algorithm 4, the $\textbf{Pool}_{non}$ consists of all $2^7=128$ treatment
	combinations, and the particles come from randomly choosing 8 of them. We set $S = 100$, $q_{GB} = 2$, $q_{LB} = 2$ , $q_{NEW} = 4$ and then get the minimum aberration design $d_2^\ast$ with $\mathcal{G}_1$-$\text{MA}=\{0, 0, 7, 7, 0, 0, 1\}$. To assess our method, we check the the following Plackett-Burman design $\mathbf{PB_8}$ given by Plackett and Burman (1946) \cite{PB}:
	
	\[\tiny
	\mathbf{PB_8}=
	\begin{blockarray}{rrrrrrr}
	A & B  & C & D  & E & F & G\\
	\begin{block}{[rrrrrrr]}
	1&1&1&1&1&1&1 \\
	1&1&-1&-1&-1&-1&1\\
	1&-1&1&1&-1&-1&-1\\
	1&-1&-1&-1&1&1&-1\\
	-1&1&1&-1&1&-1&-1\\
	-1&1&-1&1&-1&1&-1\\
	-1&-1&1&-1&-1&1&1\\
	-1&-1&-1&1&1&-1&1\\
	\end{block}
	\end{blockarray}
	,\quad
	d_2^\ast=
	\begin{blockarray}{rrrrrrr}
	A & B  & C & D  & E & F & G\\
	\begin{block}{[rrrrrrr]}
	-1&-1&-1&1&1&1&-1\\
	-1&1&1&-1&1&-1&-1\\
	-1&-1&-1&-1&-1&-1&1\\
	1&-1&1&-1&-1&1&-1\\
	-1&1&1&1&-1&1&1\\
	1&1&-1&1&-1&-1&-1\\
	1&1&-1&-1&1&1&1\\
	1&-1&1&1&1&-1&1\\
	\end{block}
	\end{blockarray}.
	\]
	The $\mathbf{PB_8}$ has the same $\mathcal{G}_1$-$\text{MA}$ as $d_2^\ast$, while they have different treatment combinations. This is a case where the minimum aberration design is nonregular.
	
		\subsubsection*{Sixteen-run nonregular design with three-stage manufacturing processes}
	We consider a three-stage manufacturing process with 16 experimental units. 
	This is equivalent to arranging a 16-run nonregular designs into a Latin square. This design has an orthogonal block structure, rather than a simple block structure, given by $\mathfrak{B} = \{\mathcal{U},\mathcal{R}, \mathcal{C}, \mathcal{L}, \mathcal{E}\}$ with $\mathcal{E} \prec \mathcal{R}, \mathcal{C}, \mathcal{L} \prec \mathcal{U}$, where
	$\mathcal{R}, \mathcal{C},\mathcal{L}$ represent the row, column and Latin stratum, respectively. When applying Theorem 2, we need to check $\mathcal{G}_1 = \{\mathcal{U}\},\  \mathcal{G}_2 = \{\mathcal{U},\mathcal{R}\},\ \mathcal{G}_3 = \{\mathcal{U}, \mathcal{C}\},\ 
	\mathcal{G}_4 = \{\mathcal{U}, \mathcal{L}\},\ \mathcal{G}_5 = \{\mathcal{U},\mathcal{R}, \mathcal{C}\},\ \mathcal{G}_6 = \{\mathcal{U},\mathcal{R}, \mathcal{L}\},\ \mathcal{G}_7 = \{\mathcal{U}, \mathcal{C}, \mathcal{L}\}$ and $\mathcal{G}_8 =\{\mathcal{U},\mathcal{R}, \mathcal{C}, \mathcal{L}\}$. 
	
	Suppose there are 6 treatment factors and the Latin square is $4\times4$. 
	We use the Latin square that is constructed by an OA(16, $4^3$, 2) given below:
			\[\small
	\text{OA(16,$4^3$,2)}=
	\begin{bmatrix}
		0& 0& 0 \\
		2& 0& 2\\
		1& 0& 1\\
		3& 0& 3\\
		0& 2& 2\\
		2& 2& 0\\
		1& 2& 3\\
		3& 2& 1\\
		0& 1& 1\\
		2& 1& 3\\
		1& 1& 0\\
		3& 1& 2\\
		0& 3& 3\\
		2& 3& 1\\
		1& 3& 2\\
		3& 3& 0\\
	\end{bmatrix}.
	\]

	Since $\xi$s are unknown, we suggest to search for designs that have minimum aberration with respect to $\mathcal{G}_1 = \{\mathcal{U}\}$. 
	We find that there are many designs with the same $\mathcal{G}_1$-MA$=\{0, 0, 0, 3, 0, 0\}$. 
	Among them, let $d_3^\ast$ be the one constructed by the following OA(16, $2^6$, 2) and $d_4^\ast$ be the design that exchanges the first row with the ninth row of $d_3^\ast$:
	\[\small
\text{OA(16,$2^6$,2)}=
\begin{pmatrix}
	-1& -1 &-1& -1& -1& -1\\
	-1& 1 &-1 &1& -1& 1\\
	-1& -1 &1& 1& -1& -1\\
	-1& 1 &1& -1& -1& 1\\
	-1& -1 &-1& -1& 1& 1\\
	-1& 1& -1& 1& 1& -1\\
	-1& -1& 1& 1& 1& 1\\
	-1& 1& 1& -1& 1& -1\\
	1& 1& 1 &1& 1& 1\\
	1& -1& 1& -1& 1& -1\\
	1& 1&-1 &-1& 1& 1\\
	1& -1& -1& 1& 1& -1\\
	1& 1& 1 &1& -1& -1\\
	1& -1& 1& -1& -1& 1\\
	1& 1& -1 &-1& -1& -1\\
	1& -1& -1& 1& -1& 1
\end{pmatrix}.
\]
	 The performances of $d_3^\ast$ and $d_4^\ast$ are given in Table 4. We can see that $d_3^\ast$ performs well in $\mathcal{G}_4, \mathcal{G}_6$ and $\mathcal{G}_8$, while $d_4^\ast$ performs well in $\mathcal{G}_3$ and $\mathcal{G}_5$. 
	\begin{table}[h]  
		\centering  
		\caption{Minimum aberration designs with three-stage manufacturing processes} 
		\label{tab5}
		\begin{tabular}{l|l|l}  
			\hline
			& $d_3^\ast$ &  $d_4^\ast$  \\ [0.5ex] 
			\hline
			$\mathcal{G}_1$-MA & \{0, 0, 0, 3, 0, 0\}    &  \{0, 0, 0, 3, 0, 0\}\\
			$\mathcal{G}_2$-MA & \{0, 7, 0, 7, 0, 1\}    &  \{0, 7, 0, 7, 0, 1\} \\
			$\mathcal{G}_3$-MA & \{2, 2, 4, 5, 2, 0\}    & \{1.75, 2, 4.5, 5, 1.75, 0\} \\
			$\mathcal{G}_4$-MA & \{1, 2, 6, 5, 1, 0\}   &   \{1.25, 2, 5.5, 5, 1.25, 0\}\\
			$\mathcal{G}_5$-MA & \{2, 9, 4, 9, 2, 1\}    &  \{1.75, 9, 4.5, 9, 1.75, 1\}\\
			$\mathcal{G}_6$-MA & \{1, 9, 6, 9, 1, 1\}    & \{1.25, 9, 5.5, 9, 1.25, 1\} \\
			$\mathcal{G}_7$-MA & \{3, 4, 10, 7, 3, 0\}    & \{3, 4, 10, 7, 3, 0\} \\
			$\mathcal{G}_8$-MA & \{3, 11, 10, 11, 3, 1\}    & \{3.25, 11, 9.5, 11, 3.25, 1\} \\
			\hline	
		\end{tabular}
	\end{table}

\subsubsection*{Fish patty experiment}
Goos (2022) \cite{Goos(2022)} revisited a fish patty experiment provided in Cornell (1981) and Cornell and Gorman (1984) \cite{CornellGorman(1984)} for strip-plot analysis, involving seven mixtures of three fish species (mullet, sheepshead and croaker) and eight processing conditions defined by an oven cooking temperature (375 or 425 degrees Fahrenheit), an oven cooking time (25 or 40 minutes) and a deep fat frying time (25 or 40 seconds).
We denote the three mixture factors by $x_1, x_2, x_3$, each having two levels coded as $-1$ and 1, where the treatment combination $(x_1, x_2, x_3)=(-1,-1,-1)$ will not be considered since it corresponds to zero percentage of the three factors.
The three two-level processing factors are denoted by $z_1, z_2, z_3$ with coded values $-1$ and 1.
Thus, there are total $(2^3-1)\times 2^3=56$ combinations in the experiment.
Goos (2022) \cite{Goos(2022)} argued that this experiment can be viewed as a strip-plot experiment.
While Goos (2022) \cite{Goos(2022)} focused on the data analysis aspect, we consider a design issue through the use of the proposed algorithm.

When the experimental cost is limited, it would be necessary to conduct a fractional factorial design instead of a full factorial design.
We use Algorithm 4 to search for a 28-run minimum aberration strip-plot fractional factorial design by halving the run size of the three processing factors.

The block structure for a strip-plot design can be represented by $\mathfrak{B} = \{\mathcal{U},\mathcal{R}, \mathcal{C}, \mathcal{E}\}$ with $\mathcal{E} \prec \mathcal{R}, \mathcal{C} \prec \mathcal{U}$, where
$\mathcal{R}, \mathcal{C}$ represent the row and column stratum, respectively.
By Theorem 2, we need to consider four aberration criteria corresponding to $\mathcal{G}_1 = \{\mathcal{U}\},\  \mathcal{G}_2 = \{\mathcal{U},\mathcal{R}\},\ \mathcal{G}_3 = \{\mathcal{U}, \mathcal{C}\},\ 
\mathcal{G}_4 = \{\mathcal{U},\mathcal{R}, \mathcal{C}\}$.
Algorithm 4 found a design that has minimum aberration with respect to $\mathcal{G}_i$-MA in 10 seconds, $i=1,2,3,4$, with the following wordlength patterns:
\begin{align*}
&\mathcal{G}_1\text{-MA}: \{0.05357, 0.05357, 0.89286, 0.05357, 0.05357, 0.01786\}, \\
&\mathcal{G}_2\text{-MA}: \{2.67857, 2.83928, 1.21429, 0.26786, 0.10714, 0.01786\}, \\
&\mathcal{G}_3\text{-MA}: \{2.625, 2.625, 1.750, 2.625, 2.625, 0.875\}, \\
&\mathcal{G}_4\text{-MA}: \{5.25000, 5.41071, 2.07143, 2.83929, 2.67857, 0.87500\}.
\end{align*}
By Theorem 2, this design has minimum aberration with respect to (\ref{Wd}) for all feasible $\pmb{\xi}$.
In addition, we find that this minimum aberration design has the treatment defining relation $I=x_1x_2x_3$, which meets the intuition.

	\section{Conclusion}
	This paper aims at developing an efficient SIB algorithm for searching for good designs with respect to the aberration criterion given by Chang and Cheng (2018) \cite{Chang2018}. We propose a novel idea which treats defining words in regular designs as particles in the SIB algorithm. We also modify the original SIB algorithm to have more sources of particles. With the assistant of design key templates, the proposed method is able to reduce computational time in searching for minimum aberration designs compared to a complete search over all possible defining words.
	In the case of nonregular designs, we use the complete factorial as the candidates of particles. When the number of treatment factors is large, the computation burden is heavy.
	Thus a direction of future research is to apply some well-structured nonregular designs such as the quaternary-code designs proposed in Phoa and Xu (2009) \cite{12} to further reduce the computational time. 
	
	
	\section*{Appendix}
	\subsection*{Design key}
	Design keys (Chapter 7 in Cheng (2014) \cite{10}) can describe the relation between the treatment factors
	and the unit factors by an $n\times2^n$ matrix $\mathbf{Y}$ such that for each $j$, $1\leq j\leq2^n$, the $j$th column of $\mathbf{Y}$ is the level combination of the unit factors corresponding to the $j$th unit. We also need an $n\times2^n$ matrix $\mathbf{X}$ whose
	$j$th column gives the treatment combination assigned to the $j$th unit. The
	matrix $\mathbf{X}$ then produces the design. Hence,
	\begin{align*}
	\mathbf{X}=\mathbf{K}\mathbf{Y},\text{ where $\mathbf{K}$ is called a design key matrix.}
	\end{align*}
	For example, in a blocked factorial design, $2^n$ treatment combinations
	are to be assigned to $2^n$ units partitioned into $2^b$ blocks of size $2^{n-b}$. For
	ease of construction, we further consider each level of $\mathcal{B}$ as a combination
	of $b$ pseudo (unit) factors $\mathcal{B}_1,\dots,\mathcal{B}_b$ with 2-level each, and each level
	of $\mathcal{P}$ as a combination of ${n-b}$ pseudo (unit) factors $\mathcal{P}_1,\dots,\mathcal{P}_{n-b}$,  also
	with 2-level. Then a main-effect or interaction contrast of the $n$ factors
	$\mathcal{B}_1,\dots,\mathcal{B}_b,\mathcal{P}_1,\dots,\mathcal{P}_{n-b}$ represents an interblock (respectively, intrablock)
	contrast if and only if it involves none (respectively, at least one) of the
	$\mathcal{P}_{j}$'s. The design key requires that the treatment combination $(x_1,\dots,x_n)^T$
	assigned to the experimental unit $(p_1,\dots,p_{n-b},b_1,\dots,b_b)^T$ satisfies
	\begin{align*}
	x_i=\sum_{j=1}^{n-b}k_{i,j}p_{j}+\sum_{l=1}^{b}k_{i,n-b+l}b_{l}.
	\end{align*}	
	The equation  indicates that under the constructed design, the main effect
	of the $i$th treatment factor coincides with the ``factorial'' effect of the
	unit factors defined by $(k_{i,1},\dots,k_{i,n-b},k_{i,n-b+1},\dots,k_{i,n})$, the $i$th row of $\mathbf{K}$. We say that the latter is the unit alias of the former. The main effect
	of the $i$th treatment factor is estimated in the intrablock stratum if its
	unit alias involves at least one of the $\mathcal{P}_{j}$'s, which means that at least one
	of $(k_{i,1},\dots,k_{i,n-b}$ is nonzero. In order to generate all the $2^n$ treatment
	combinations, the $n$ columns of $\mathbf{K}$ must be linearly independent. Further, one can determine the treatment interactions that are confounded
	with blocks by identifying those whose unit aliases involve $\mathcal{B}_1,\dots,\mathcal{B}_b$
	only. This information can also be obtained from the inverse relation $\mathbf{Y}=\mathbf{K}^{-1}\mathbf{X}$. In $\mathbf{K}^{-1}$, which is called the \textit{inverse design key matrix}, the roles of treatment and unit factors are reversed, with its rows corresponding
	to unit factors and its columns corresponding to treatment factors. Thus the rows of $\mathbf{K}^{-1}$ determine the treatment factorial effects that are
	confounded with main effects of the unit factors and can be used to determine
	the treatment factorial effects that are estimated in each stratum.\\
	
		\textbf{Lemma 1} (Cheng (2014) \cite{10}) Suppose neither the row nor the column
	design is a replicated fraction. Then each additional basic factor for the
	row or column design must be defined by a word of the form XY, where
	both X and Y involve at least one factor, and X (respectively, Y) involves
	the row (respectively, column) treatment factors in the basic design key
	only. If XY defines a row (respectively, column) treatment factor, then Y
	(respectively, X) is a blocking word.\\
	
	\textbf{Theorem 3}\quad (Cheng and Tsai (2013) \cite{Designkey}) Suppose $\mathfrak{B}_1$ and $\mathfrak{B}_2$ are simple block structures on $s^{m_1}$ and $s^{m_2}$ units, respectively, where $s$ is a
	prime number or power of a prime number, and $n = m_1+m_2$. Then, subject
	to factor relabeling, a complete $s^n$ factorial design with block structure
	$\mathfrak{B}_1/\mathfrak{B}_2$ or $\mathfrak{B}_1\times \mathfrak{B}_2$ can be constructed by using a design key of the form
	\begin{align*}
	\mathbf{K}=\left[ \begin{array}{cc}
	\mathbf{K}_2 & \mathbf{A}_2\\
	\mathbf{A}_1 & \mathbf{K}_1\\
	\end{array}
	\right]
	\end{align*}
	where $\mathbf{K}_1$ and $\mathbf{K}_2$ are design keys for complete $s^{m_1}$ and $s^{m_2}$ factorial designs
	with block structures $\mathfrak{B}_1$ and $\mathfrak{B}_2$, respectively, and $\mathbf{A}_1$ and $\mathbf{A}_2$ are some
	matrices, with the first $m_2$ columns of $\mathbf{K}$ corresponding to the unit factors
	in $\mathfrak{B}_2$ and the last $m_1$ columns corresponding to the unit factors in $\mathfrak{B}_1$. $\mathbf{A}_2$
	is used to describe the situation of crossing. If there is no crossing, than
	$\mathbf{A}_2=\mathbf{0}$.\\
	
	\textbf{Remark 1}\quad  Under complete 2-level factorial designs, each row in $\mathbf{K}^{-1}$ is a
	defining word and $\mathbf{K}^{-1}=\mathbf{K}$.

	\subsection*{PSO and SIB}
	The PSO is a population-space-based optimization tool proposed by Kennedy
	and Eberhart (1995) \cite{11}. The PSO has been shown in many high-dimensional
	optimization problems that it usually converges or nearly converges to the
	global optimum. The PSO is commented that it has some advantages in terms
	of flexible forms of optimum and constraints, small number of tuning parameters
	and minimal memory space for mathematical operations. The standard PSO algorithm begins by first randomly generating a swarm
	of particles to search for the optimal solution. The size of the swarm is
	user-selected and each particle in the swarm represents a candidate solution
	to the optimization problem. At any particular time, each particle
	has its own perceived location of the optimum in the search space and
	with each iteration. This perceived optimum location by the particle at
	a given iteration is often called the ``Local Best'' (LB). Another position
	called ``Global Best'' (GB) comes from all particles sharing information on
	the perceived overall optimum location. For the $i$th particle after completing
	its $n$th iteration, the following two equations show how the new $i$th
	particle $x_{i,n+1}$ is determined and what velocity $v_{i,n+1}$ to use to set there:
	\begin{align*}
	v_{i,n+1}&=v_{i,n}+C_1(x_{GB}-x_{i,n})+C_2(x_{i,LB}-x_{i,n})\\
	x_{i,n+1}&=x_{i,n}+v_{i,n+1}\times\delta_t.
	\end{align*}
	Here $v_{i,j}$ and $x_{i,j}$ are respectively the velocity and position of particle $i$ at
	iteration $j$, $C_2$ and $C_2$ are the stochastic terms, $x_{GB}$ is the GB position as perceived by all particles
	and $x_{i,LB}$ is the LB position for
	particle $i$. The new location for the next iteration is the sum of its current
	location and the velocity it uses to get there at unit time when $\delta_t = 1$.
	
	\textit{Swarm intelligence-based} (SIB) in Phoa (2016) \cite{SIB} is an algorithm
	for solving discrete optimization problems. The SIB, which inherits the characteristic
	of particle referencing to GB and LB in the PSO, can be viewed as a
	discrete version of the PSO. A moving mode cannot be described with velocity
	vector in the SIB. The author proposed two operations \textit{MIX} and \textit{MOVE} to
	describe a new moving mode.
	
	\begin{algorithm}[htb] 
		\caption{SIB algorithm} 
		\label{alg:Framwork} 
		\begin{algorithmic}[1] 
			\STATE Randomly generate a set of initial particles.
			\STATE Evaluate objective function value of each particle.
			\STATE Initialize the LB for all particles.
			\STATE Initialize the GB.
			\STATE \textbf{While} not converge \textbf{do}.
			\STATE For each particle, perform the MIX operation.
			\STATE For each particle, perform the MOVE operation.
			\STATE Evaluate objective function value of each particle.
			\STATE Update the LB for all particles and update the GB.
			\STATE \textbf{End while}.
		\end{algorithmic}
	\end{algorithm}
	
	The MIX operation considers a candidate particle X and another particle Y which is either the LB or the GB particles. The general idea is to improve
	the values of objection functions of X via substituting $q$ (user-defining)
	``bad'' units of X by $q$ ``good'' units of Y. $q$ represents $q_{LB}$ or $q_{GB}$ when Y
	is the LB or the GB particles respectively. First, deletion step for X is a
	recursive procedure that determines which unit is the best unit to removed
	from X, hence, each $q$ will recursive total units in a particle. In one of $q$,
	if $X_{-r}$ which consists of all units of X except the $r$th unit has the best
	objective function value. It deletes the $r$th unit from X in the current step and
	treat the reduced particle as X. Deletion step continues sequentially until $q$
	units are deleted from the original X. Second, addition step is the reverse of
	the deletion step, where $q$ units from Y are sequentially added in a similar
	way to reducing X. We call new X a \textit{mixwGB} or \textit{mixwLB}, respectively Y
	is LB or GB.
	
	The MOVE operation comes after the MIX operations of all X are completed
	and there are three candidates, mixwGB, mixwLB and X, under
	consideration. The MOVE operation is a decision-making procedure to
	select the best particles among all candidates with the optimal value of objective
	functions. Among them if both mixwGB and mixwLB are worse
	than X, some units of X randomly chosen and replaced by some other random
	units.

	%
	%



\end{document}